\documentclass[12pt]{article}
\usepackage{a4wide}
\usepackage{graphicx}
\usepackage{epsfig}

\def\ba{\begin{eqnarray}}
\def\ea{\end{eqnarray}}
\def\beq{\begin{equation}}
\def\eeq{\end{equation}}
\def\bas{\begin{eqnarray*}}
\def\eas{\end{eqnarray*}}

\def\imag{{\rm i}}
\def\rhobar{\overline \rho}
\def\etabar{\overline \eta}

\begin{document}

\begin{center}
\Large{ \bf 
Quark-lepton complementarity with lepton and quark mixing data predict
$\theta_{13}^{PMNS}=(9{^{+1}_{-2}})^{\circ}$}
\end{center}

\begin{center}
Bhag C. Chauhan$^{a\star}$\footnote{On leave from Govt. Degree College, Karsog (H P) India 171304.}, 
Marco Picariello$^{b\star}$, 
Jo\~{a}o Pulido$^{a\star}$,
Emilio Torrente-Lujan$^{c\star}$\\[2mm]
{\small\sl
$^a$ 
Centro de F\'{\i}sica Te\'{o}rica das Part\'{\i}culas (CFTP)
Departamento de F\'{\i}sica,
\\Instituto Superior T\'{e}cnico
Av. Rovisco Pais, P-1049-001 Lisboa, Portugal\\
$^b$
Dipartimento di Fisica, Universit\`a di Lecce and INFN-Lecce,\\
Via Arnesano, ex Collegio Fiorini, I-73100 Lecce, Italia\\
$^c$
Dep. de Fisica, Grupo de Fisica Teorica,
Univ. de Murcia,  Murcia, Spain.
}
\end{center}

\begin{abstract}
\noindent
The complementarity between the quark and lepton
 mixing matrices is shown to provide a robust prediction for
 the neutrino mixing angle $\theta_{13}^{PMNS}$. We obtain this prediction by 
 first showing that the matrix $V_M$, product of the CKM and PMNS mixing 
 matrices, may have a zero (1,3) entry which is 
 favored by experimental data. Hence models with bimaximal or
 tribimaximal forms of the correlation matrix $V_M$ are quite possible.
 Any theoretical model with a vanishing (1,3) entry of $V_M$ that is
 in agreement with quark data, solar, and atmospheric
 mixing angle leads to 
 $\theta_{13}^{PMNS}=(9{^{+1}_{-2}})^\circ$. This value is consistent
 with the present 90\% CL experimental upper limit. 


\end{abstract}

\vfill
{\small \noindent $\star$
chauhan@cftp.ist.utl.pt, 
Marco.Picariello@cern.ch,
pulido@cftp.ist.utl.pt,
torrente@cern.ch}

\newpage

\section{Introduction}

Recent neutrino experiments confirm that 
the Pontecorvo-Maki-Nakagawa-Sakata
(PMNS) \cite{Pontecorvo:1967fh,Maki:1962mu}
lepton mixing matrix $U_{PMNS}$ contains large mixing angles.
For example the atmospheric mixing $\theta_{23}^{PMNS}$
is compatible with $45^\circ$~\cite{Fogli:2005gs},
and the solar mixing $\theta_{12}^{PMNS}$ is
$\approx 34^\circ$~\cite{Aliani:2003ns}.
These results should be compared with the third lepton mixing angle
$\theta_{13}^{PMNS}$ which is very small and even compatible with zero 
\cite{Apollonio:1999ae,Aliani:2002na}, and with the quark mixing angles 
in the $U_{CKM}$ matrix~\cite{Cabibbo:1963yz,Kobayashi:1973fv}.

The disparity that nature indicates between quark and lepton mixing
angles has been viewed in terms of a 'quark-lepton complementarity' (QLC)
\cite{Minakata:2004xt} which can be expressed in the relations
\beq
\theta_{12}^{PMNS}+\theta_{12}^{CKM}\simeq 45^\circ\,;
\quad\quad
 \theta_{23}^{PMNS}+\theta_{23}^{CKM}\simeq 45^\circ\,.
\eeq

Possible consequences
of QLC have been investigated in the literature~\cite{Ferrandis:2004mq}
and in particular a simple correspondence between the $U_{PMNS}$ and $U_{CKM}$
matrices has been proposed~\cite{Minakata:2004xt,
Frampton:2004vw,Antusch:2005ca,Ma:2005qy}
and analyzed in terms of a 
correlation matrix~\cite{Xing:2005ur,Dighe:2006zk,
Rodejohann:2003sc,Frampton:2004ud,Datta:2003qg,Datta:2005ci,Harada:2005km}.
The correlation matrix $V_M$ is simply defined as the product 
of the CKM and PMNS matrices,
$V_M=U_{CKM}\cdot  U_{PMNS}$, and efforts
have been done to obtain the {\em most favorite} pattern
for the matrix $V_M$~\cite{Harada:2005km,Raidal:2004iw}. 
Unitarity then implies $U_{PMNS}=U^{\dagger}_{CKM}V_M$ 
and one may ask where do the large lepton mixings come from?
Is this information implicit in the form of the $V_M$ matrix?
This question has been widely investigated in the literature,
but its answer is still open (see our section {\bf 2}).

Furthermore in some Grand Unification Theories (GUTs) the direct 
QLC correlation between the $CKM$ and the $PMNS$ mixing matrix can
be obtained. In this class of models, the $V_M$ matrix is determined
by the heavy Majorana neutrino mass
matrix~\cite{Antusch:2005ca,Georgi:1979df}.
Moreover as long as quarks and leptons are inserted in the same 
representation of the underlying gauge group,
we need to include in our definition of $V_M$ arbitrary
but non trivial phases between the quark and lepton matrices. 
Hence we will generalize the relation $V_M=U_{CKM}\cdot  U_{PMNS}$
to 
\begin{equation}\label{eq:fund}
V_M=U_{CKM}\cdot \Omega \cdot U_{PMNS}\,
\end{equation}
where the quantity $\Omega$ is a diagonal matrix
$\Omega={\rm diag}(e^{\imag\omega_i})$ and the three phases
$\omega_i$ are free parameters (in the sense that they are
not restricted by present experimental evidence).

The magnitude disparity between the lepton mixing angle $\theta_{13}^{PMNS}$
and the other two mixings is a rather striking fact.
In this paper we carry out the investigation of the correlation matrix $V_M$ 
based on eq.~(\ref{eq:fund}) and prove that it is a zero texture of $V_M$, namely
$V_{M_{13}}=0$, that implies a small value for $\theta_{13}^{PMNS}$ with 
a sharp prediction
\begin{equation}\label{eq:prediction}
\theta_{13}^{PMNS}=(9\pm^{1}_{2})^{\circ}.
\end{equation} 
We use the Wolfenstein parameterization for $U_{CKM}$ 
\cite{Wolfenstein:1983yz} in its unitary form~\cite{Buras:1994ec}
and parameterize $U_{PMNS}$ with the standard
phases and mixing angles. 
As a zero order approximation we start inserting by hand the central values 
of the lepton mixing angles and CKM parameters. 
Owing to the uncertainty in the experimental value for $\theta_{13}^{PMNS}$,
the possible range for the (1,3) entry of matrix $V_M$
may or may not include zero.
For example using $\theta_{13}^{PMNS}=3^\circ$ the (1,3) entry range does
not include zero in accordance with eq.~(23) in ref.~\cite{Xing:2005ur}. For 
other choices of $\theta_{13}^{PMNS}$ a vanishing (1,3) entry is quite 
possible, as will be seen in section~{\bf\ref{sec:VM}}.

It is possible to include bimaximal and tribimaximal  
forms of the correlation matrix $V_M$ in models with 
renormalization effects~\cite{Kang:2005as,Cheung:2005gq,Ellis:1999my}
that are relevant, however, only in particular cases with large 
$\tan\beta$ $(> 40)$ and with quasi degenerate 
neutrino masses~\cite{Antusch:2005gp}.
The conclusion for matrix $V_M$ is that the possibility of a 
bimaximal form, or a tribimaximal one is completely open.
So in other words, the correlation between the matrices $U_{CKM}$ and 
$U_{PMNS}$ is rather nontrivial.

The investigation we perform for the $V_M$ matrix starts from the 
fundamental equation $V_M=U_{CKM}\cdot \Omega \cdot U_{PMNS}$ and
uses the experimental ranges and constraints on lepton mixing angles. 
We resort to a Monte Carlo simulation with two-sided Gaussian distributions
around the mean values of the observables. 
The input information on $\theta_{13}^{PMNS}$ is taken from the analysis of
ref.~\cite{Fogli:2005gs} which uses neutrino data only.

The paper is organized as follows: 
in section~{\bf\ref{sec:VM}} we study the numerical ranges of $V_M$ entries 
with the aid of a Monte Carlo simulation, emphasizing on specific points of 
the experimental data. We will show that the vanishing of the $(1,3)$ entry 
is favored by the data analysis.
In section~{\bf\ref{sec:PMNS}} we present the matter from a different
point of view:
we start from a zero $(1,3)$ $V_M$ entry (e.g. a bimaximal or 
tribimaximal matrix) we derive the consequent prediction for the $U_{PMNS}$ 
lepton mixing matrix through 
\beq
U_{PMNS}=(U_{CKM}\cdot  \Omega)^{-1}\cdot  V_M
\eeq
and the corresponding one for $\theta_{13}^{PMNS}$ in
eq.~(\ref{eq:prediction}).
Finally we present a summary and our conclusions.

\section{Which $V_M$ does the phenomenology imply?}\label{sec:VM}

In this section we investigate the order of magnitude of the $V_M$ matrix
entries concentrating in particular in the (1,3) entry and the important 
mixing angle $\theta_{13}^{V_{M}}$ to which it is directly related. We then
explicitly study the allowed values of the $V_M$ angles, finally concluding 
that $sin^2 \theta_{13}^{V_{M}}=0$ is the value most favored by the data.
We will be using the Wolfenstein parameterization \cite{Wolfenstein:1983yz}
of the $U_{CKM}$ matrix in its unitary form \cite{Buras:1994ec}
where one has the relation 
\beq\label{CKM}
\sin \theta_{12}^{CKM}=\lambda
\quad\quad
\sin \theta_{23}^{CKM}=A\lambda^2
\quad\quad
\sin \theta_{13}^{CKM}e^{-\imag \delta^{CKM}}=A \lambda^3(\rho-\imag \eta)
\eeq
to all orders in $\lambda$.
The lepton mixing matrix $U_{PMNS}$ is parameterized in the usual way as
\ba
U_{PMNS}=
 U_{23}\cdot \Phi\cdot U_{13}\cdot\Phi^\dagger\cdot U_{12}\cdot \Phi_m.
\ea
Here $\Phi$ and $\Phi_m$ are diagonal matrices containing the Dirac and
Majorana CP violating phases, respectively
$\Phi={\rm diag}(1, 1, e^{\imag \phi})$
and $\Phi_m={\rm diag}(e^{\imag \phi_1}, e^{\imag \phi_2}, 1)$, so that
\ba\label{PMNS}
U_{PMNS}=\pmatrix{
e^{\imag\,\phi_1} c_{12}\,c_{13} &
            e^{\imag\,\phi_2} c_{13}\,s_{12} &
                   s_{13}e^{-\imag \,\phi}\cr
 e^{\imag\,\phi_1}\left(
    -c_{23}\,s_{12}-e^{\imag \,\phi}\,c_{12}\,s_{13}\,s_{23}\right)&
e^{\imag\,\phi_2}\left(
            c_{12}\,c_{23} - e^{\imag \,\phi}\,s_{12}\,s_{13}\,s_{23}\right)&
                   c_{13}\,s_{23}\cr
 e^{\imag\,\phi_1}\left(
   - e^{\imag \,\phi}\,c_{12}\,c_{23}\,s_{13} + s_{12}\,s_{23}\right)&
e^{\imag\,\phi_1}\left(
            - e^{\imag \,\phi}\,c_{23}\,s_{12}\,s_{13}-c_{12}\,s_{23}\right)&
                   c_{13}\,c_{23}\,
}
\ea

\subsection{Estimation of $V_M$ entries}\label{naive}

In grand unification models where quarks and leptons belong to the same 
representation of the gauge group, the quark and lepton fields must
acquire different phases once their symmetry is broken. Hence one
should take into account this phase mismatch at low energy associated
with the form of the CKM and PMNS matrices (\ref{CKM},~\ref{PMNS}).
To this end we introduced the diagonal matrix $\Omega$
\ba\label{Omega}
\Omega={\rm diag}(e^{\imag \omega_1},e^{\imag \omega_2},e^{\imag\omega_3})\,
\ea
in the commonly used relation\footnote{see e.g. refs. \cite{Minakata:2004xt,
Xing:2005ur}.}
$V_M=U_{CKM} \cdot U_{PMNS}$. This is therefore generalized to
\ba\label{eq:ThVM}
V_M&=&U_{CKM}\cdot \Omega \cdot U_{PMNS}\,.
\ea
We use for the observed CKM mixing parameters the values
$\lambda=0.2237$, $\eta=0.317$, $\rho=0.225$,
$|V_{cb}|\approx A \lambda^2=0.041$,
and for the PMNS mixing angles the values
$\theta_{12}^{PMNS}=34^\circ$, $\theta_{23}^{PMNS}=45^\circ$,
$\theta_{13}^{PMNS}=3^\circ$~\cite{Xing:2005ur}. 
For the $\Omega$ phases we resort
a Monte Carlo simulation with flat distributions in the 
interval $[0,2\pi]$. We then get the following range of values for 
the elements of the $V_M$ correlation matrix:
\ba\label{eq:xing}
V_M=\left(
\begin{array}{ccc}
0.71...0.91 & 
0.41...0.68 & 
0.10...0.22 
\\
0.15...0.62 & 
0.40...0.74 & 
0.65...0.75 
\\
0.34...0.45 & 
0.54...0.64 & 
0.68...0.72
\end{array}
\right).
\ea
These values are in good agreement with~\cite{Xing:2005ur}.
The small differences are due to the fact that we use the
full mixing matrix given in eq.~(\ref{CKM}) and not the 
parameterization given in eq.~(21) of Ref.~\cite{Xing:2005ur}.
Notice that the $(1,3)$ entry of the matrix $V_M$
above cannot be zero, so $V_M$ cannot be bimaximal, i.e.
of the form
\beq
\pmatrix{
\frac{1}{\sqrt{2}}&\frac{1}{\sqrt{2}}&0\cr
\frac{1}{2}&\frac{1}{2}&\frac{1}{\sqrt{2}}\cr
\frac{1}{2}&\frac{1}{2}&\frac{1}{\sqrt{2}}
}
\simeq
\pmatrix{
0.71 & 0.71 & 0.00\cr
0.50 & 0.50 & 0.71\cr
0.50 & 0.50 & 0.71
}\,,
\eeq
nor tribimaximal, namely
\beq
\pmatrix{
\sqrt{\frac{2}{3}}&\frac{1}{\sqrt{3}}&0\cr
\frac{1}{\sqrt{6}}&\frac{1}{\sqrt{3}}&\frac{1}{\sqrt{2}}\cr
\frac{1}{\sqrt{6}}&\frac{1}{\sqrt{3}}&\frac{1}{\sqrt{2}}
}
\simeq
\pmatrix{
0.82& 0.58& 0.00\cr
0.41& 0.58& 0.71\cr
0.41& 0.58& 0.71
}\,,
\eeq
where only the absolute values have been considered.
The result of eq.~(\ref{eq:xing}) however depends on the assumption about the
values used for the mixing angles.
For example if we use a different value for $\theta_{13}^{PMNS}$, namely 
$\theta_{13}^{PMNS}=9.2^\circ$ (see Ref.~\cite{Fogli:2005gs} or our
eq.~(\ref{bestfit2})
for the allowed range of $\theta_{13}^{PMNS}$), we get
\beq
V_M=\left(
\begin{array}{ccc}
0.69...0.88 & 
0.39...0.67 & 
0.00...0.32
\\
0.09...0.67 & 
0.36...0.78 & 
0.62...0.75 
\\
0.28...0.51 & 
0.49...0.68 & 
0.67...0.73 
\end{array}
\right)\,.
\eeq
For these values the result is in agreement with the statement
that $V_M$ has the $(1,3)$ entry equal to zero.
It is clear that we need a better investigation
of the situation before establishing what are the allowed values of the
entries of the correlation matrix $V_M$ that can be deduced from
the experimental data.
We next investigate the important entry $(1,3)$ as it overwhelmingly affects 
the $\theta_{13}^{PMNS}$ prediction as will be seen in
section~{\bf\ref{sec:PMNS}}.

We parameterize the $V_M$ correlation matrix as the
$PMNS$ lepton mixing matrix, i.e.
\ba\label{VMpar}
V_M\equiv
 U_{23}\cdot \Phi\cdot U_{13}\cdot\Phi^\dagger\cdot U_{12}
\ea
where $U_{ij}$s are functions of the mixing angles $\theta_{ij}^{V_M}$.

\begin{figure}
\centering
{\epsfig{file=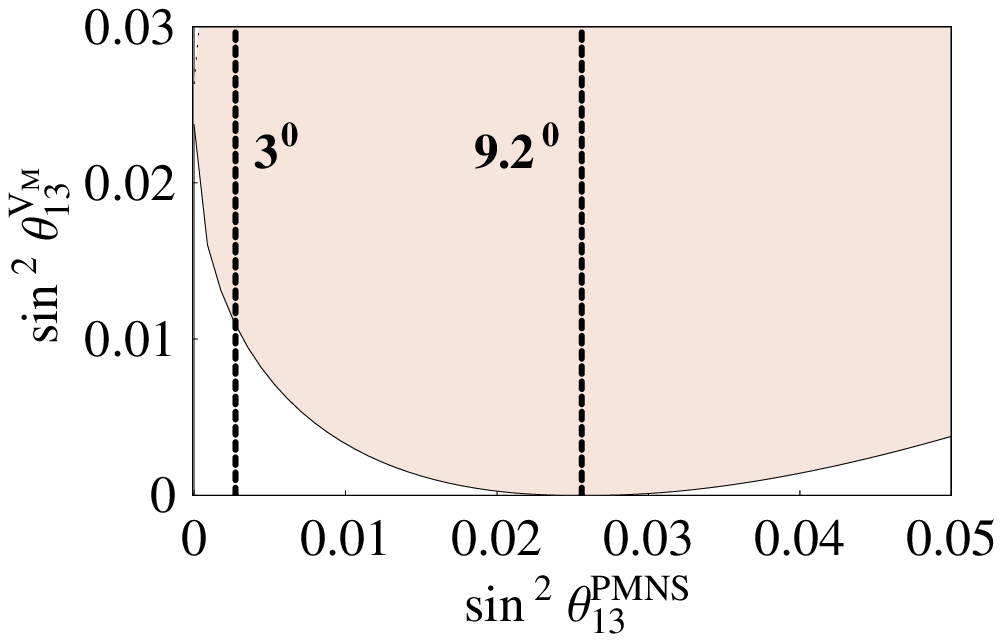,height=6.cm}}
\caption{The minimum
value allowed for $\sin^2\theta_{13}^{V_M}$
as a function of $\sin^2\theta_{13}^{PMNS}$.
All the other CKM and PMNS mixing parameters are fixed at their best
fit points given in eq.~(\ref{bestfit}-\ref{bestfit2}).
The unknown phases $\omega_1$,  $\omega_2$, and $\omega_3$
of $\Omega$, the Majorana phases $\phi_1$, and $\phi_2$, and the
Dirac one $\phi$  
are taken to vary within the interval $[0,2\,\pi]$ with a flat
distribution.
We also report the values of $\theta_{13}^{PMNS}=3.0^\circ$ and $9.2^\circ$
used in the text.
}
\label{fig:f1}
\end{figure}

At first non trivial orders
in $\lambda$ we have
\ba\label{th13VM_A}
\sin^2\theta_{13}^{V_M}&=&
\left|\left(1-\frac{\lambda^2}{2}\right) e^{\imag (\omega_1 -\omega_2-\phi)}
\sin \theta_{13}^{PMNS}+\lambda \sin \theta_{23}^{PMNS} \cos \theta_{13}^{PMNS}
+O(\lambda^3)\right|^2
\ea
It is seen from this expression that the first two terms can cancel each other
implying a vanishing $(1,3)$ entry of the $V_M$ matrix.
In order to better investigate this issue we plot in fig.\ref{fig:f1}
the quantity $\sin^2\theta_{13}^{V_M}$
 as a function of $\sin^2\theta_{13}^{PMNS}$.
All other observables are fixed at their best fit points
\cite{Fogli:2005gs,Aliani:2003ns,Charles:2004jd} and we allowed
the Dirac lepton phase $\phi$, the Majorana ones
$\phi_1$ and $\phi_2$, and the unphysical phases of $\Omega$
to vary in the interval $[0,2\,\pi]$ with a flat distribution.

As shown in the figure, for the central value of $\theta_{13}^{PMNS}$ given 
in~\cite{Fogli:2005gs} the entry (1,3) of $V_M$ cannot be zero.
However there is a small region $(\theta_{13}^{PMNS}\approx 9.2^\circ)$ 
for which $\theta_{13}^{V_M}$ can be zero.
This fact has the very important consequence of providing a sharp
prediction for the unknown mixing angle $\theta_{13}^{PMNS}$.
We will investigate this point in detail in the section~{\bf\ref{sec:PMNS}}.

\subsection{The allowed values for $\tan^2\theta_{23}^{V_M}$,
 $\tan^2\theta_{12}^{V_M}$, and $\sin^2\theta_{13}^{V_M}$}\label{sec:thetaVM}

Here we further investigate the possibility of $V_M$ to be bimaximal or
tribimaximal using the fundamental equation~(\ref{eq:ThVM}). We start with 
a Monte Carlo simulation for the $U^{CKM}$ parameters, the $U^{PMNS}$ 
mixing angles, the $\Omega$ and CP phases. 

We use the updated values for the $CKM$ and $PMNS$ mixing matrix,
given at $95\%$CL by \cite{Charles:2004jd}
\ba\label{bestfit}
\begin{tabular}{cc}
$\lambda = 0.2265^{+0.0040}_{-0.0041}$
\,,&
$A=0.801^{+0.066}_{-0.041}$
\,,\cr\cr
$\etabar = 0.189^{+0.182}_{-0.114}$
\,,&
$\rhobar = 0.358^{+0.086}_{-0.085}$
\,,
\end{tabular}
\ea
with 
\beq
\rho + i\eta =
\frac{\sqrt{1-A^2\lambda^4}(\rhobar+i\etabar)}{\sqrt{1-\lambda^2}
\left[1 - A^2\lambda^4(\rhobar+i\etabar)\right]}\,;
\eeq
and~\footnote{The lower uncertainty for $\sin^2\theta_{13}$ is purely formal,
and correspond to the positivity constraint $\sin^2\theta_{13}\geq 0$.} 
\cite{Fogli:2005gs,Aliani:2003ns}
\ba\label{bestfit2}
\begin{tabular}{cc}
$\sin^2\theta_{23}^{PMNS} = 0.44\times\left(1^{+0.41}_{-0.22}\right)$
\,,&
$\sin^2\theta_{12}^{PMNS} = 0.314\times\left(1^{+0.18}_{-0.15}\right)$
\,,\cr\cr
$\sin^2\theta_{13}^{PMNS} = \left(0.9^{+2.3}_{-0.9}\right)\times 10^{-2}$
\,.
\end{tabular}
\ea
With the aid of a MonteCarlo program we generated the values for
each variable: for the sine square of the lepton mixing angles and
for the quark parameters $A,~\lambda,~\bar\rho,~\bar\eta$ we took
two sided Gaussian distributions with central values and standard
deviations taken from eqs. (\ref{bestfit}-\ref{bestfit2}). For the unknown phases we
took flat random distributions in the interval $[0,2\pi]$. We divided
each variable range into short bins and counted the number of occurences
in each bin for all the variables, having run the program $10^6$ times.
In this way the corresponding histogram is smooth and the number of
occurences in each bin is identified with the probability density
at that particular value. A comparatively high value of this
probability density extending over a wide range in
the variable domain means a high probability for the variable to lie
in this range, therefore that such range is 'favoured' by the
data being used as MonteCarlo input. Conversely higher probability
implies better compatibility with experimental data, while lower
probability means poor or no compatibility with data.

\begin{figure}
\centering
{\epsfig{file=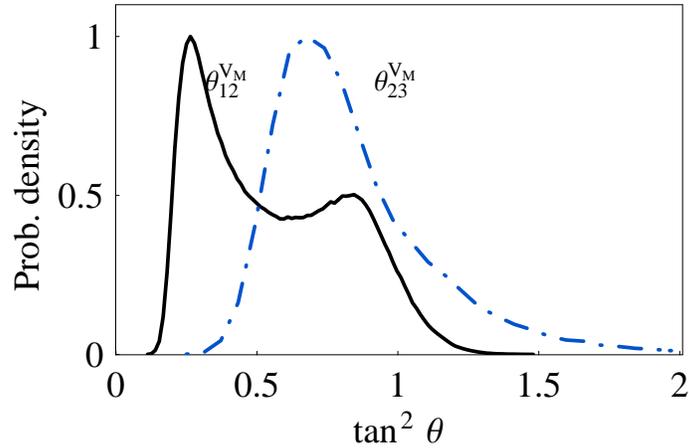,height=6.cm}} 
\caption{
The distributions, normalized to one at the maximum,
of $\tan^2\theta_{12}^{V_M}$ (solid),
and $\tan^2\theta_{23}^{V_M}$ (dot-dashed)
obtained from the definition of the correlation mixing matrix $V_M$
given in eq.~(\ref{eq:ThVM}) by using a Monte Carlo simulation
of all the experimental data.}
\label{fig:f2}
\end{figure}

\begin{figure}
\centering
{\epsfig{file=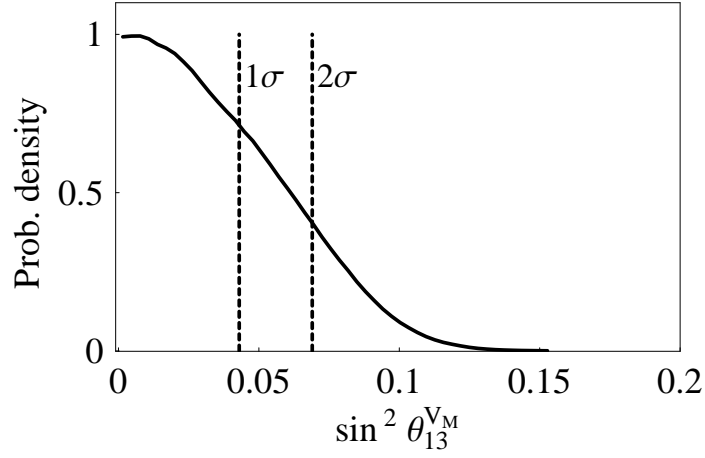,height=6.cm}}
\caption{The distribution, normalized to one at the maximum,
of $\sin^2\theta_{13}^{V_M}$
 obtained from the definition of the correlation mixing matrix $V_M$
given in eq.~(\ref{eq:ThVM}) by using a Monte Carlo simulation
of all the experimental data.
We also plot the $1\sigma$ and the $2\sigma$ lines.}
\label{fig:f3}
\end{figure}

In figs~\ref{fig:f2} and \ref{fig:f3}
we report the results of this simulation.
The distributions of $\tan^2\theta_{23}^{V_M}$ and 
$\tan^2\theta_{12}^{V_M}$ are shown in fig \ref{fig:f2}.
It is seen that the range for which the value of
$\tan^2\theta_{23}^{V_M}$ is compatible with experiments
at $90\%$CL is the interval $[0.35, 1.4]$, so that 
$\tan^2\theta_{23}^{V_M}=1.0$ is consistent with data.
For $\tan^2\theta_{12}^{V_M}$ we obtain a range between
$0.25$ and $1.1$ at $90\%$CL and so $\tan^2\theta_{12}^{V_M}=1.0$ 
(which corresponds to a bimaximal matrix) only
within 3$\sigma$.
Moreover the value $\tan^2\theta_{12}^{V_M}=0.5$ (which corresponds
to a tribimaximal matrix), is well inside the allowed range.
Finally in fig.\ref{fig:f3} we plot the distribution for
$\sin^2\theta_{13}^{V_M}$.
We see that $\sin^2\theta_{13}^{V_M}=0$
is not only allowed by the experimental data,
 but also it is the preferred value.
In the next section we will see that this has important consequences
in the model building of flavor physics.

\section{Prediction for $\theta_{13}^{PMNS}$}\label{sec:PMNS}

In this section
we investigate the consequences of a $V_M$ correlation matrix
with zero (1,3) entry on the still experimentally undetermined
$\theta_{13}^{PMNS}$ mixing angle.
In particular we will see that the $\theta_{13}^{PMNS}$
prediction arising from eq.~(\ref{eq:ThVM}) or, equivalently,
\ba\label{PMNSexp}
U_{PMNS}&=&(U_{CKM}\cdot \Omega)^{-1}\cdot V_M
\ea
is quite stable against variations in the form of $V_M$ allowed
by the data.

As previously shown (see section~{\bf\ref{sec:thetaVM}}),
the data favours a vanishing (1,3) entry in $V_M$.
So in the whole following analysis we fix
$\sin^2\theta_{13}^{V_M}=0$. We allow the 
$U_{CKM}$ parameters to vary, with a two-sided Gaussian distribution,
within the experimental ranges given in eq.~(\ref{bestfit}), while for the 
$\Omega$ phases in eq. (\ref{Omega}) we take flat distributions in the interval
$[0,2\pi]$.

We make Monte Carlo simulations for different values of $\theta_{12}^{V_M}$ 
and $\theta_{23}^{V_M}$ mixing angles, allowing
$\tan^2\theta_{12}^{V_M}$ and $\tan^2\theta_{23}^{V_M}$ 
to vary respectively within the intervals $[0.3,1.0]$ and $[0.5,1.4]$ in 
consistency with the lepton and quark mixing angles
(see section~{\bf\ref{sec:thetaVM}} and fig.~\ref{fig:f2}).

\begin{figure}
\centering
\begin{tabular}{cc}
{\epsfig{file=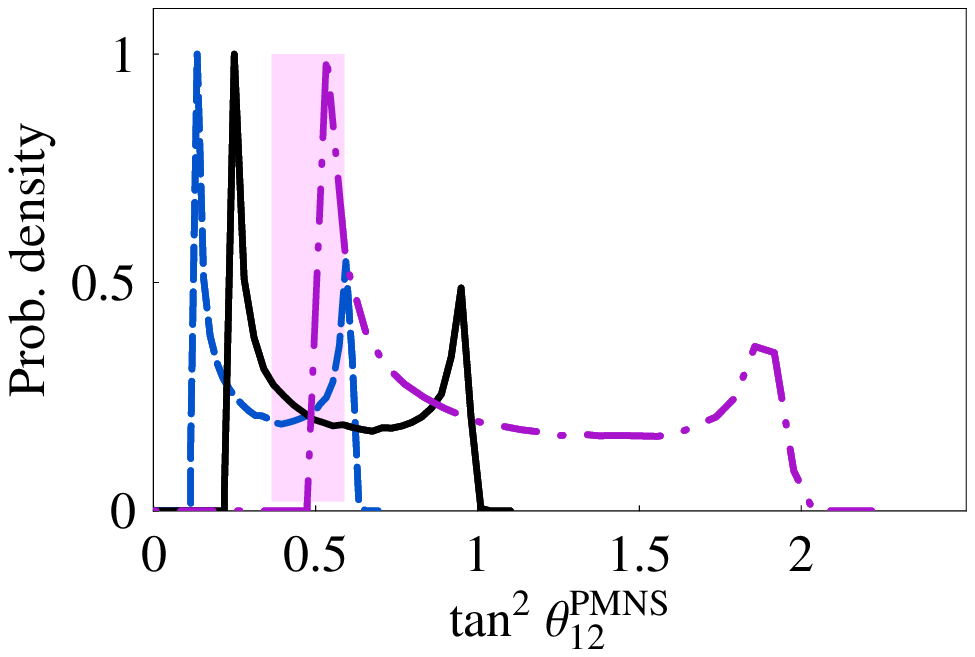,height=4.75cm}} &
{\epsfig{file=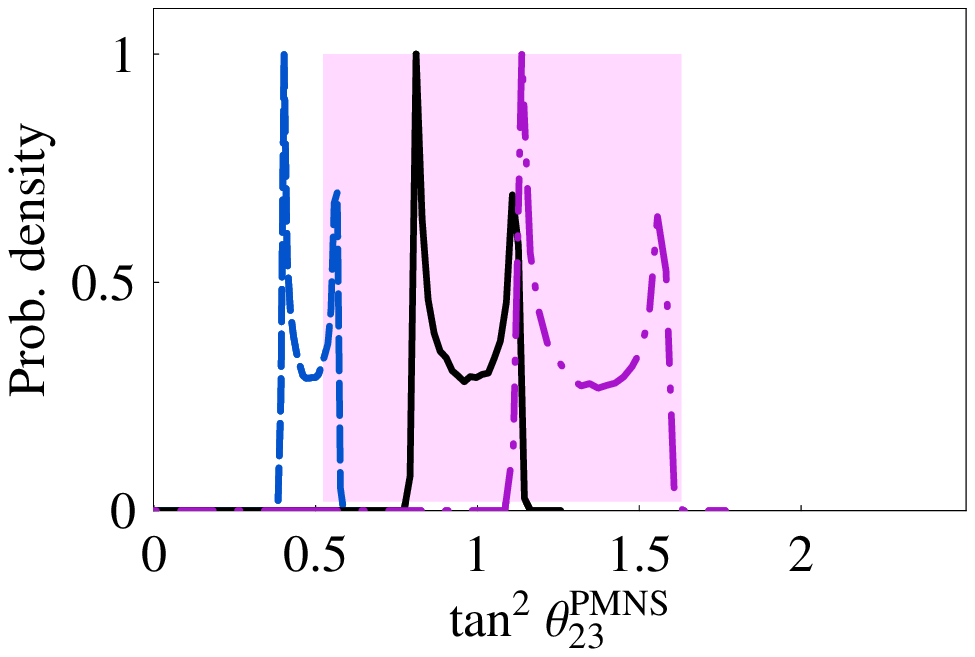,height=4.75cm}}
\end{tabular}
\caption{
The distribution of $\tan^2\theta_{12}^{PMNS}$ (left),
and $\tan^2\theta_{23}^{PMNS}$ (right)
for the CKM experimental data and
for values of the correlation matrix $V_M$ respectively given by
(left) $\tan^2\theta_{12}^{V_M}=
0.3$ (dashed),
$0.5$ (solid),
$1.0$ (dot-dashed),
$\tan^2\theta_{23}^{V_M}=1.0$,
and $\sin^2\theta_{13}^{V_M}=0$; 
(right)
$\tan^2\theta_{23}^{V_M}=
0.5$ (dashed),
$1.0$ (solid),
$1.4$ (dot-dashed), $\tan^2\theta_{12}^{V_M}=0.5$,
$\sin^2\theta_{13}^{V_M}=0$.
The shaded areas represent
the experimentally allowed regions at $2 \sigma$ for each case.}
\label{fig:f4a}
\label{fig:f4b}
\end{figure}

In fig.\ref{fig:f4a}(left) we plot the distribution of
$\tan^2\theta_{12}^{PMNS}$
for values of the correlation matrix $V_M$ corresponding to 
$\tan^2\theta_{12}^{V_M}\in\{0.3,0.5,1.0\}$ with $\tan^2\theta_{23}^{V_M}=1.0$.
From the figure we can check that for $\tan^2\theta_{12}^{V_M}=0.3,~{\rm and}~0.5$ 
the resulting distribution for
$\tan^2\theta_{12}^{PMNS}$ is compatible with the experimental data. 
Instead maximal $\theta_{12}^{V_M}$ and $\theta_{23}^{V_M}$ taken together
are disfavoured, as the solar angle is hardly compatible
with the corresponding allowed interval (dot-dashed line).

In fig.\ref{fig:f4b}(right)
 we plot the distribution of $\tan^2\theta_{23}^{PMNS}$
for $\tan^2\theta_{23}^{V_M}\in\{0.5,1.0,1.4\}$ with
$\tan^2\theta_{12}^{V_M}=0.5$.
Also in these cases we see that the resulting distributions for
$\tan^2\theta_{23}^{PMNS}$ are compatible with the experimental data.

\begin{figure}
\centering
{\epsfig{file=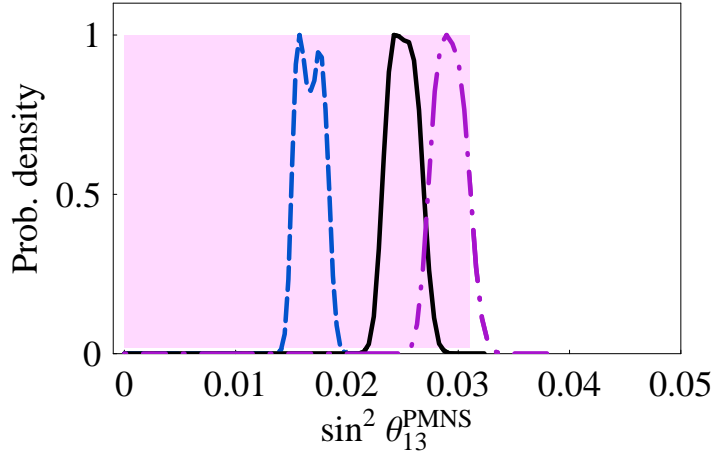,height=6.cm}} 
\caption{
The distribution of $\sin^2\theta_{13}^{PMNS}$
for the CKM experimental data and
for values of the correlation matrix $V_M$ given by
$\tan^2\theta_{12}^{V_M}=0.5$, $\sin^2\theta_{13}^{V_M}=0$,
$\tan^2\theta_{23}^{V_M}=
0.5$ (dashed),
$1.0$ (solid),
$1.4$ (dot-dashed).
The shaded area represents
the experimentally allowed region at $2 \sigma$.}
\label{fig:f4c}
\end{figure}

Finally we report in fig.\ref{fig:f4c} the results of our simulation for
the quantity $\sin^2\theta_{13}^{PMNS}$.
From eq.~(\ref{PMNSexp}), the parameterization of the CKM mixing matrix
in eq.~(\ref{CKM}) and the definition of the phase matrix $\Omega$ in
eq.~(\ref{Omega}) we get  
\ba
(U_{PMNS})_{13} &=&
e^{-i\omega_1}\Big[
\left(1-\frac{\lambda^2}{2}\right)\sin \theta_{13}^{V_M}e^{-i \phi^{V_M}}
-\lambda \sin \theta_{23}^{V_M} \cos \theta_{13}^{V_M}
\nonumber\\&&\quad\quad\quad
 + A\lambda^3(-\rho+i\,\eta+1)\cos \theta_{23}^{V_M} \cos \theta_{13}^{V_M}
+ O(\lambda^4)
 \Big]\,,
\ea
so that
\ba\label{th13PMNS_A}
\sin^2\theta_{13}^{PMNS} &=&
\sin^2\theta_{23}^{V_M}\lambda^2 + O(\lambda^3)\,,
\ea
where we have used the fact that $\sin^2\theta_{13}^{V_M}=0$
and $A\approx O(1)$.

From eq.~(\ref{PMNSexp}) and the parameterization used for $V_M$
in eq.~(\ref{VMpar})
we see that $\sin^2\theta_{13}^{PMNS}$ does
 not depend on $\tan^2\theta_{12}^{V_M}$.
For this reason the parameter $\sin^2\theta_{13}^{PMNS}$ needs to be studied
as a function of $\tan^2\theta_{23}^{V_M}$ only. Fixing for definiteness
$\tan^2\theta_{12}^{V_M}=0.5$ and taking the three different values
$\tan^2\theta_{23}^{V_M}\in\{0.5, 1.0, 1.4\}$, we plot in fig.6  
the corresponding distributions of $\sin^2\theta_{13}^{PMNS}$. 
We note that these values of $\tan^2\theta_{23}^{V_M}$ practically 
cover the whole range consistent with the data (see fig.2).

From fig.~\ref{fig:f4c}
 it is seen that the $\sin^2\theta_{13}^{PMNS}$ distributions are
quite sharply peaked around maxima of $7.3^\circ$, $8.9^\circ$ and 
$9.8^\circ$. Recalling that the shift of this maximum is effectively
determined 
by the parameter $\tan^2\theta_{23}^{V_M}$ which was chosen to span most
of its physically allowed range, it is clear that we have a stable 
prediction for $\theta_{13}^{PMNS}$.

\begin{figure}
\centering
{\epsfig{file=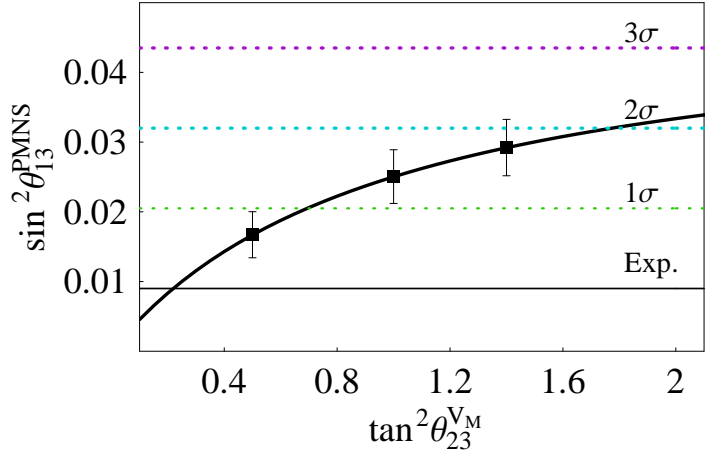,height=6.cm}} 
\caption{The allowed values for $\sin^2\theta_{13}^{PMNS}$ as a function of 
$\tan^2\theta_{23}^{V_M}$ under the assumption that
 $\sin^2\theta_{13}^{V_M}=0$.
We report the central and $3 \sigma$ values obtained from fig.\ref{fig:f4c},
and the approximate analytical dependence given in eq.~(\ref{th13PMNS_A}).
We also plot the experimental central value, the $1\sigma$, the $2\sigma$,
and the $3\sigma$ from~\cite{Fogli:2005gs}. 
We fixed $\tan^2\theta_{12}^{V_M}=0.5$ for definiteness.}
\label{fig:f5}
\end{figure}

In order to better clarify this stability, we show 
in fig.~\ref{fig:f5} the mean and the standard deviation of 
$\sin^2\theta_{13}^{PMNS}$ obtained with our Monte Carlo simulation for the 
three chosen values of $\tan^2\theta_{23}^{V_M}$.
In addition we plot the analytic
dependence of $\sin^2\theta_{13}^{PMNS}$ given by eq.~(\ref{th13PMNS_A}) with 
the central value of $\lambda$, the best fit point of
$\sin^2\theta_{13}^{PMNS}$ and its 1$\sigma$, 2$\sigma$
and 3$\sigma$ from the analysis of ref.\cite{Fogli:2005gs}. 
Our prediction for $\theta_{13}^{PMNS}$ then follows from the experimental data on
$\lambda$ ,$A$, $\rho$, $\eta$, $\tan^2\theta_{12}^{PMNS}$
and $\tan^2\theta_{23}^{PMNS}$ and the values of $\tan^2\theta_{12}^{V_M}$,
$\tan^2\theta_{23}^{V_M}$ are taken in the intervals $[0.3,1.0]$, $[0.5,1.4]$ 
respectively, as allowed by the data. For a vanishing $(1,3)$ entry of the matrix 
$V_M$ we finally find $\theta_{13}^{PMNS}$ in the interval $[7^\circ,10^\circ]$.  

To conclude this section we note that another prediction for a small 
$\theta_{13}^{PMNS}$ has recently been derived \cite{Harada:2005km} 
\begin{equation}
\theta_{13}^{PMNS}=9^{\circ}+O(sin^{3}\theta_{12}^{CKM}).
\end{equation}
This follows from an assumed bimaximality of a matrix relating
Dirac to Majorana neutrino states together with the assumption 
that neutrino mixing is described by the CKM matrix at the grand unification
scale. Our approach on the other hand is free from any {\it ad hoc} assumptions. 
We show that it is a zero texture of the $V_M$
correlation matrix, namely $V_{M_{13}}=0$, together with all the experimental values
of the quark and lepton mixing angles, that predicts
$\theta_{13}^{PMNS}=(9 \pm^{1}_{2})^{\circ}$. More importantly we show that the
vanishing of this entry is favored by the data. 
Condition $V_{M_{13}}=0$ is compatible with $V_M$ being 
bimaximal (i.e. with two angles of $45^\circ$ and a
vanishing one), tribimaximal (i.e. with one angle of $45^\circ$, one with
$\tan^2\theta=0.5$ and a third vanishing one) or of any other form.
Furthermore we make use of a phase matrix $\Omega$, see 
eqs.~(\ref{Omega}-\ref{eq:ThVM}), that takes account of the mismatch between 
the quark and lepton phases
and consider Majorana phases in the $U_{PMNS}$ matrix with a flat random 
distribution. 

\section{Summary and Conclusions}\label{conclusion}

In summary, we have investigated the correlation between the CKM quark and 
PMNS lepton mixing matrices, arising in a large class of GUT seesaw models
with specific flavor symmetries.
The detailed analysis developed here uses the fact that the correlation
matrix is phenomenologically compatible with a tribimaximal pattern,
and marginally with a bimaximal pattern.
This conclusion is different from the one obtained in 
previous studies \cite{Xing:2005ur} and is in 
agreement with other qualitative arguments that favor the  
CKM matrix to measure the deviation of the PMNS matrix from 
exact bimaximal mixing~\cite{Raidal:2004iw}.

In our analysis we found that the mixing parameters 
$\tan^2\theta_{12}^{V_M}$ and $\tan^2\theta_{23}^{V_M}$ vary respectively 
within the intervals $[0.3,1.0]$ and $[0.5,1.4]$, while 
$\sin^2\theta_{13}^{V_M}$ varies in the range $[0.0, 0.2]$. Moreover the 
preferred value for $\sin^2\theta_{13}^{V_M}$ is zero.

Using these results we investigated the phenomenological
consequences of correlation matrices $V_M$
with zero $(1,3)$ entry.
The main conclusion of this study is that this large class
of models is not only compatible with the experimental data, but
also that they give a robust prediction for $\theta_{13}^{PMNS}$
mixing angle 
\begin{equation}\label{eq:last}
\theta_{13}^{PMNS}= (9{^{+1}_{-2}})^{\circ}\,.
\end{equation}
Whereas the author of ref.\cite{Harada:2005km} obtains a 
prediction for $\theta_{13}^{PMNS}$ in a similar range, our 
result cannot be regarded as a straightforward extension or generalization.
In fact the condition $V_{M_{13}}=0$, which
is favored by the data, is the only requirement for the prediction
(\ref{eq:last}).
 Furthermore we modified the correlation between the CKM and PMNS
mixing matrices to take account of a phase matrix $\Omega$ between
the quark and lepton fields. Eq.~(\ref{eq:last})
 will be checked with great accuracy
in the next generation of precision neutrino experiments (DCHOOZ and others).

We studied GUT models with flavor symmetry that predict a relation of 
the type $V_{M}=U_{CKM} \cdot \Omega \cdot U_{PMNS}$ with $V_{M_{13}}=0$.
Since in supersymmetric models with $\tan\beta \leq 40$ radiative corrections
are small \cite{Kang:2005as,Cheung:2005gq,Ellis:1999my,Antusch:2005gp},
this relation can in such cases be used at 
low energy as in the present paper. Hence if future dedicated
experiments exclude $\theta_{13}^{PMNS}\simeq 9^{\circ}$ and supersymmetry is
discovered with $\tan\beta \leq 40$, such models would be ruled out.
On the other hand, a positive result from $\theta_{13}^{PMNS}$ dedicated experiments
and $\tan\beta \leq 40$ would be a strong hint for these flavor symmetry models 
and its specific Higgs pattern.

\subsection*{Acknowledgments}

The work of BCC was supported by Funda\c{c}\~{a}o para a
Ci\^{e}ncia e a Tecnologia through the grant SFRH/BPD/5719/2001.
We acknowledge the MEC-INFN grant,
Fundacion Seneca (Murcia) grant,
CYCIT-Ministerio of Educacion (Spain) grant. M.P. and
E.T-L. would like to thank Milan University for kind hospitality.
M.P. would like to thank the Instituto Superior T\'{e}cnico for
kind hospitality.

\newpage


\end{document}